\documentclass[prb,twocolumn,amsmath,amssymb]{revtex4-2}

\usepackage{graphicx}
\usepackage{dcolumn}
\usepackage{bm}
\usepackage{mathrsfs}
\usepackage{subfigure}
\usepackage{natbib}
\usepackage{amsmath}
\usepackage{amssymb}
\usepackage{Jonasmacros}
\usepackage{mathptmx}
\usepackage{txfonts}
\usepackage{ifsym}
\usepackage[usenames,dvipsnames,svgnames]{xcolor}

\usepackage[colorlinks,citecolor=magenta,linkcolor=red]{hyperref}
\usepackage[usenames,dvipsnames,svgnames]{xcolor}

\begin{document}
\date{\today}

\title{Emergent magnetism as a cooperative effect of interactions and reservoir}

\author{M. Shiranzaei}
\author{S. Kalh\"ofer}
\author{J. Fransson}
\email{Jonas.Fransson@physics.uu.se}
\affiliation{Department of Physics and Astronomy, Box 516, 751 20, Uppsala University, Uppsala, Sweden}


\begin{abstract}
Closed shell molecular structures are under normal conditions time-reversal invariant. Experimental evidences point, however, towards that this invariance may be locally violated when the structure is in contact with a particle reservoir. The mechanisms behind such local symmetry breaking are not clear by any means. By considering a minimal model for a closed shell structure, here we propose that the symmetry breaking may result from a combination of internal and/or external interactions. It is shown that a magnetic moment of a localized electron level can be generated and maintained under the influence of such combination. The theoretical results should cast new light on the mechanisms that may form magnetic properties in molecular compounds.
\end{abstract}
\maketitle

Molecules, as well as single atoms, which are in closed shell configurations when isolated can acquire a magnetic state when, e.g., immersed in solution \cite{ChemSci.1.631,Science.331.445}, attached on a surface \cite{Nature.468.1476,NatPhysics.8.497,JMaterChemC.3.11986,Science.352.318,RSCAdv.9.34421,PhysRevMaterials.5.114801,ACSNano.16.13049}, or being in embedded in clusters comprising several components \cite{JACS.125.8694,NatComms.6.10139,ChemSci.7.6132,JPhysChemLett.7.4988,JPhysChemC.121.12159,ApplPhysExpress.13.113001,ACSNano14.16624,JPhysChemC.125.9875}. Such properties can be exploited in, for instance, anomalous Hall devices \cite{NatComms.8.14567,JPhysChemLett.10.1139,ApplPhysLett.118.172401}, electron spin resonance \cite{Science.350.417,ChemSci.13.12208}, exploration of superconductivity in presence of spin impurities \cite{NatPhysics.9.765} giving rise to Yu-Shiba-Rusinov states \cite{Science.332.940,NanoLett.19.5167}, and in structures with properties, such as coercivity \cite{JPhysChemLett.7.4988,JPhysChemC.121.12159,ApplPhysExpress.13.113001,ACSNano14.16624,JPhysChemC.125.9875} and spin-filtering \cite{JACS.142.17572,ACSNano14.16624,JPhysChemC.126.3257,JACS.144.7302}, that strengthens with temperature.

The origin of the magnetic state in the closed shell configuration can be effectively summarized as an interplay between the Pauli exclusion principle and the Hund's rules. Although these rules with some success can be employed also in a more general context, questions about the emergence of magnetic states in molecules that are normally regarded as non-magnetic inevitably arise. For instance, chiral molecules provide urgent examples of closed shell structures which, nevertheless, display magnetic properties when in contact with otherwise non-magnetic metals, see, e.g., Refs. \citenum{NatComms.8.14567,Small.1801249,JPhysChemLett.10.1139,PhysRevMaterials.5.114801,ChemSci.13.12208,JACS.144.7302}.

In this article we address the issue of the emergence of a magnetic state in or in a proximity around a local electronic structure when it is being exposed to an external environment. We begin by demonstrating that a spin degenerate molecular level may become spin-polarized if two conditions are met. First, there should exist internal molecular interactions which have the potential to break the time-reversal symmetry and second, the molecular level must be in contact with an external reservoir. We show that the nature of the reservoir, whether it is Fermionic or Bosonic, is secondary. This observation, hence, implies that also molecules in a purely thermal environment may be spontaneously polarized.

As a corollary result of these conditions, we also show that the spin-degeneracy of a localized electron may be broken by a spin-dependent coupling to a purely Bosonic reservoir. Breaking of the spin-degeneracy requires, however, the presence of both spin-conserving \emph{and} spin-nonconserving coupling. In this model we, furthermore, demonstrate the emergence of a non-vanishing magnetic moment and an associated cross-over temperature at which this moment undergoes a sign change.

We explain our findings to be a result of confluent interactions since these results cannot be obtained in a system with a single type of interaction. For simplicity, assume that there are two sources of interactions which can be formulated through the quantities $V_1\sigma^0$ and $\bfV_1\cdot\bfsigma$, where $\sigma^0$ and $\bfsigma$ are the $2\times2$-unit matrix and vector of Pauli spin matrices. When these two interaction coexist, the effective interaction changes the spectrum as $(V_0\sigma^0+\bfV_1\cdot\bfsigma)^2=(V_0^2+|\bfV_1|^2)\sigma_0+2V_0\bfV_1\cdot\bfsigma$, which opens for the possibility to break the spin-degeneracy whenever both $V_0$ and $\bfV_1$ contributes.

As a philosophical remark, our results are important since they challenge the wide spread view that we can interpret measurements in terms of subsystems where the environment has a negligible effect, and we present a concrete example where this is not the case. Despite that we are taught in our scientific training that a measurement inevitably influences the properties of the sample, both interpretations of experimental results as well as theoretical descriptions are  many times based on complete negligence of the reservoir to which the sample is connected.

The purpose here is to evaluate the magnetic moment $\av{\bfm_0}$ of a localized electron represented by the spectrum $\bfepsilon=\dote{0}\sigma^0+\boldsymbol{\epsilon}_1 \cdot \bfsigma$, where $\dote{0}$ and $\boldsymbol{\epsilon}_1$ denote the energies corresponding to the spin-independent and spin-dependent degrees of freedom. Here, the latter is a three component vector, $\boldsymbol{\epsilon}_1=\dote{\alpha}\hat{\bf e}_\alpha$, in some normalized orthogonal basis $\{\hat{\bf e}_\alpha\}$, which accounts for, e.g, spin-orbit interactions and local spin-anisotropy. The model corresponding to this spectrum can be written $\Hamil_0=\psi^\dagger\bfepsilon\psi$, where $\psi=(\psi_\up,\ \psi_\down)^t$ denotes the spinor for the localized state.

In order to enable a general treatment of the local properties, we calculate the expectation of the magnetic moment $\av{\bfm}$ in terms of the Green function $\bfG_\text{LS}$ for the local electron through the relation $\av{\bfm}=(-i){\rm sp}\bfsigma\int\bfG^<_\text{LS}(\omega)d\omega/4\pi$, where $\bfG^<_\text{LS}$ denotes the lesser form of the Green function, whereas ${\rm sp}$ is the trace over spin 1/2 space. The equation of motion for $\bfG_\text{LS}$ can be cast in the Dyson-like form
\begin{align}
\bfG_\text{LS}=&
	\bfg_\text{LS}
	+
	\bfg_\text{LS}\bfSigma\bfG_\text{LS}
	,
\label{eq-Dyson}
\end{align}
where $\bfg_\text{LS}=\bfg_\text{LS}(z)=(z-\bfepsilon)^{-1}$, $z\in\mathbb{C}$, is the bare Green function defined by $\Hamil_0$, whereas $\bfSigma$ denotes the self-energy caused by the interactions the local electron is subject to. In this context, one can notice that the self-energy has (i) an energy dependence, $\bfSigma=\bfSigma(z)$, and (ii) can be written on the form $\bfSigma=\Sigma_0\sigma^0+\bfSigma_1\cdot\bfsigma$, which are natural conditions for spin 1/2 particles. Physically, this partitioning represents the charge- ($\Sigma_0$) and spin-dependent ($\bfSigma_1$) components of the interactions. However, in addition we shall make the replacement $\Sigma_0\rightarrow V+\Sigma_0$. In this construction, $V$ may define a contribution caused by hybridization between the localized state and an external reservoir, whereas the self-energy $\bfSigma$ may be attributed to internal external, interactions associated with the localized electron. There is, nevertheless, nothing that prevents the opposite association of $V$ and $\bfSigma$, that is, that the former belongs to the molecule and the latter represents the interactions with the environment, as we shall see in the concrete example below.

Summarizing these facts, it is straight forward to write the retarded/advanced Green function as
\begin{align}
\bfG^{r/a}_\text{LS}(\omega)=&
	\frac{
		\Bigl(\omega-\dote{0}-V^{r/a}-\Sigma_0^{r/a}\Bigr)\sigma^0
		+
		\Bigl(\bfepsilon_1+\bfSigma_1^{r/a}\Bigr)\cdot\bfsigma
	}
	{(\omega-E^{r/a}_+)(\omega-E^{r/a}_-)}
	,
\end{align}
with the poles
\begin{align}
E^{r/a}_\pm=&
		\dote{0}+V^{r/a}+\Sigma_0^{r/a}
		\pm
		\sqrt{\Bigl(\bfepsilon_1+\bfSigma_1^{r/a}\Bigr)\cdot\Bigl(\bfepsilon_1+\bfSigma_1^{r/a}\Bigr)}
	.
\label{eq-poles}
\end{align}
Under equilibrium conditions, the fluctuation-dissipation theorem implies that the lesser Green function $\bfG_\text{LS}^<$ can be expressed in terms of its retarded counterpart, $\bfG^r_\text{LS}$, using the identity $\bfG^<_\text{LS}(\omega)= \mathrm{i} f(\omega)[-2\im\bfG_\text{LS}^r(\omega)]$, where $f(\omega)$ is the Fermi-Dirac distribution function which relates to the chemical potential $\mu$ of the system.
Of particular interest here is the component comprising the Pauli matrices, since only this term can contribute under the trace ${\rm sp}\bfsigma\bfG^<_\text{LS}$. Indeed, using the notation $\bfG_\text{LS}=G_0\sigma^0+\bfG_1\cdot\bfsigma$, it can be seen that $\av{\bfm}=(-\mathrm{i})\int\bfG^<_1(\omega)d\omega/2\pi$. 
Here,
\begin{align}
\bfG^<_1(\omega)=&
	-2 \mathrm{i} f(\omega)
	\im
	\frac{(\omega-E^a_+)(\omega-E^a_-)}
		{|\omega-E^r_+|^2|\omega-E^r_-|^2}
	\Bigl(\bfepsilon_1+\bfSigma_1^r\Bigr)
.
\end{align}

In order to sort out the origin of the induced magnetic moment, we set
\begin{subequations}
\begin{align}
\lambda=&
    \re V^r
    ,
&
\gamma=&
    -\im V^r
    ,
\\
\Lambda_0=&
    \re(\dote{0}+\Sigma^r_0)
    ,
&
\Gamma_0=&
    -\im(\dote{0}+\Sigma^r_0),
\\
\bfLambda_1=&
    \re(\bfepsilon_1+\bfSigma^r_1)
    ,
&
\bfGamma_1=&
    -\im(\bfepsilon_1+\bfSigma_1^r)
    ,
\end{align}
\end{subequations}
and keep in mind that $\Lambda_1=|\re(\bfepsilon_1+\bfSigma^r_1)|$ and, $\Gamma_1=|\im(\bfepsilon_1+\bfSigma^r_1)|$. The the lesser Green function can, then, be written
\begin{align}
\bfG^<_1(\omega)=&
	2if(\omega)
	\left\{
	\frac{(\omega-\omega_+)\Gamma_-
		+
		(\omega-\omega_-)\Gamma_+}
	{|\omega-\omega_++i\Gamma_+|^2|\omega-\omega_-+i\Gamma_-|^2}
	\bfLambda_1
	\right.
\nonumber\\&
	\left.
	+
	\frac{(\omega-\omega_+)(\omega-\omega_-)-\Gamma_+\Gamma_-}
	{|\omega-\omega_++i\Gamma_+|^2|\omega-\omega_-+i\Gamma_-|^2}
	\bfGamma_1
	\right\}
,
\end{align}
where $\omega_\pm=\lambda+\Lambda_0\pm\Lambda_1$ and $\Gamma_\pm=\gamma+\Gamma_0\pm\Gamma_1$.

As we wish to determine the origin of the magnetic moment, assume, for the sake of argument, that $\bfG^<_1$ strongly peaks at the resonance energies $\omega_\pm$, while it is nearly vanishing off resonance. This assumption is justified whenever the broadening $\Gamma_\pm$ is small in a neighborhood around $\omega_\pm$. Then, the magnetic moment can be estimated by approximately
\begin{align}
\av{\bfm}\approx&
	\frac{1}{2\pi}
	\sum_{s=\pm1}
		s
		f(\omega_s)
		\frac{\Gamma_{\bar{s}}^2}{\Gamma_s^2}
	        \frac{\Lambda_1\bfLambda_1+(\Gamma_s/2)\bfGamma_1}
		{\Lambda_1^2+(\Gamma_{\bar{s}}/2)^2}
		\bigg|_{\omega_s}
	.
\end{align}
Assuming, furthermore, that the self-energy strongly peaks at the energy $\dote{0}+\omega_0$, which does not coincide with either of $\omega_\pm$, then, one can notice that $\Gamma_0(\omega_\pm)\approx0$ and $\bfGamma_1(\omega_\pm)\approx0$, such that the magnetic moment reduces to
\begin{align}
\av{\bfm}\approx&
	\frac{1}{2\pi}
	\Biggl(
    	\frac{\Lambda_1\bfLambda_1f(\omega_+)}
    	{\gamma(\Lambda_1^2+(\gamma/2)^2)}
    	\bigg|_{\omega_+}
    	-
    	\frac{\Lambda_1\bfLambda_1f(\omega_-)}
    	{\gamma(\Lambda_1^2+(\gamma/2)^2)}
	    \bigg|_{\omega_-}
    \Biggr)
	.
\label{eq-mapprox}
\end{align}
It should be mentioned that the energy $\omega_0$ is associated with the energy of the internal interactions captured in $\bfSigma$.

Here, we stress that the parameters $\bfLambda_1=\bfLambda_1(\omega)$, $\bfGamma_1=\bfGamma_1(\omega)$, et c., and that they acquire different values at the resonances $\omega=\omega_\pm$. Hence, in the limit $\bfepsilon_1=0$, this calculation leading to Eq. \eqref{eq-mapprox} demonstrates that, despite the simplicity inferred, the result comprises a fundamentally important feature of the composite system discussed here. Namely, while the internal interactions, which lead to the self-energy $\bfSigma$, provides an energy dependent shift of the electron resonances and their corresponding life times, as well as an induced finite spin-splitting, and while the coupling between electrons in the localized level and the reservoir contributes to the level broadening of the local resonances, it is only when those two mechanisms are present simultaneously that a finite magnetic moment can be induced and maintained in the localized level.

The implications of this result should have bearing on the interpretation of experimental results, as well as, how a theoretical account for a phenomenon can be made irrelevant by exclusion of effects from the environment. In magnetism, for instance, many types of interactions which, at first sight, may appear unrelated may actually play a non-trivial role for the stabilization of the ordered state \cite{ACSNano14.16624,JPhysChemC.125.9875,VibFIM2023}. The magnetic signatures observed after adsorbing non-magnetic molecules onto metallic surface \cite{NatComms.8.14567,NanoLett.19.5167,ApplPhysLett.118.172401} stem from mechanisms that are unlikely to be captured within the conventional theory for magnetism.


It is by now established that time-reversal symmetry may be broken by inelastic scattering \cite{PhysRevB.84.201307,PhysRevB.84.201306}. Therefore, we consider a simplified example that may be used to illustrate a possible experimental outcome for single molecules in contact with a thermal reservoir. Such a system can be modeled by the Hamiltonian $\Hamil=\Hamil_\text{mol}+\Hamil_\text{ph}+\Hamil_\text{e-ph}$, where $\Hamil_\text{mol}=\psi^\dagger\bfepsilon\psi$ denotes the valence state in the molecule, whereas $\Hamil_\text{ ph } = \sum_\bfq\omega_\bfq b^\dagger_\bfq b_\bfq$ represents the thermal reservoir in which $b_\bfq^\dagger$ ($b_\bfq$) creates (annihilates) a phonon at the energy $\omega_\bfq$. The electron-phonon coupling is provided through the term
\begin{align}
\Hamil_\text{e-ph}=&
	\sum_\bfq\psi^\dag\bfU_\bfq\psi( b_\bfq+b^\dag_{\bar\bfq})
	, 
\end{align}
where the coupling parameter $\bfU_\bfq=u_{0\bfq}\sigma_0+\bfu_{1\bfq} \cdot \boldsymbol{\sigma}$, whereas $\bar\bfq=-\bfq$. In addition to $u_{0\bf q}$ which defines a generic coupling between charge and vibrational modes, $\bfu_{1\bfq}$ denotes a vibrationally induced spin-orbit coupling \cite{PhysRevB.102.235416, JPhysChemC.126.3257} . Here, $\sigma_0$ and $\boldsymbol{\sigma}$ denote the $2 \times 2$ identity and vector of Pauli matrices, respectively.

\begin{figure}[t]
\begin{center}
\includegraphics[width=\columnwidth]{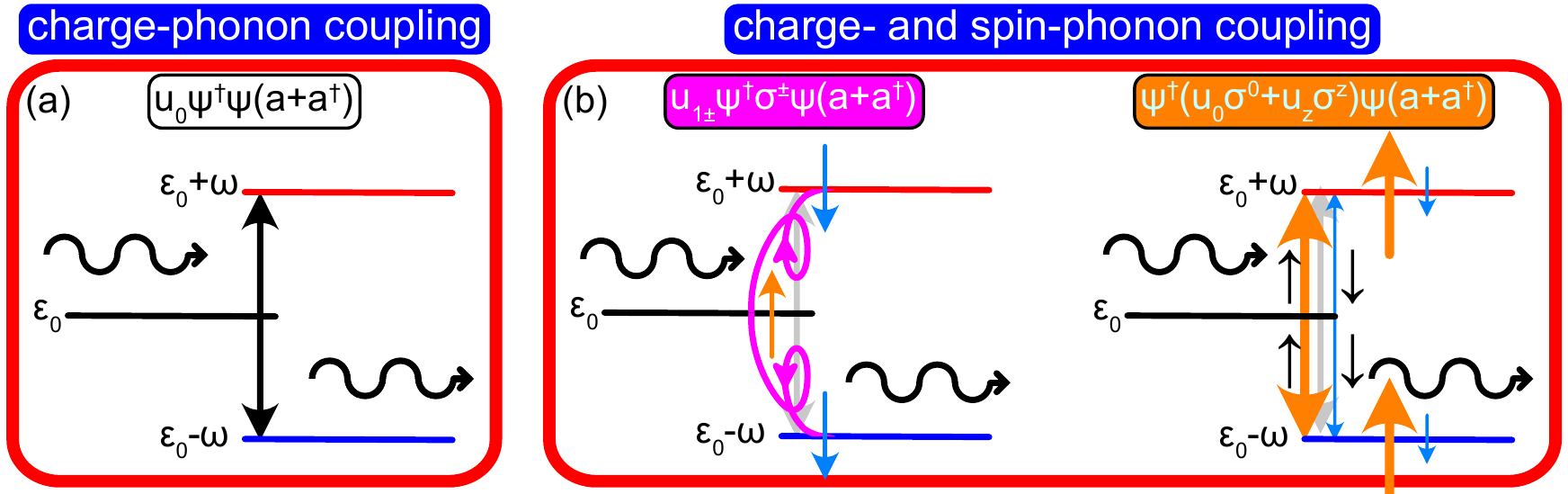}
\end{center}
\caption{Illustration of the electron-phonon processes involving (a) only charge and (b) both charge and spin.
By emission or absorption of phonons, the total charge undergoes transitions to the states at the energies $\dote{0}-\omega$ and $\dote{0}+\omega$, respectively.
(a) For a coupling solely between the charge and phonons, there is no spin related process.
(b) For a coupling that involves both charge and spin, the transitions may be accompanied by spin-flip and spin-dependent rates.
}
\label{fig-ep}
\end{figure}

The processes associated with the terms $u_{0\bfq}\psi^\dagger\psi(b_\bfq+b_{\bar\bfq}^\dagger)$ and $\psi^\dagger\bfu_{1\bfq}\cdot\bfsigma\psi(b_\bfq+b_{\bar\bfq}^\dagger)$ are illustrated in Fig. \ref{fig-ep} (a) and (b), respectively. In processes of the former kind, the electrons emit or absorb phonons such the total charge undergoes a transition to the emission or absorption state. By contrast, in processes of the latter, in which both charge and spin are coupled to the phonons, the emission and absorption processes are accompanied by electronic spin-flips and spin-dependent rates.

The magnetic moment $\av{\bfM_\text{mol}}$ is related to the lesser single electron Green function $\bfG^<_\text{mol}$, which is given by the Dyson-like equation in Eq. \eqref{eq-Dyson}. The self-energy $\bfSigma=\sum_\bfq\bfU_\bfq\tilde\bfSigma_\bfq\bfU_{\bar\bfq}$ is in the second order approximation given by the electron-phonon exchange loop \cite{PhysRevB.102.235416},
\begin{align}
\tilde\bfSigma(z)=&
	\frac{1}{\beta}
	\sum_\nu
		\bfG_\text{mol}(z-z_\nu)
		D_\bfq(z_\nu)
	,
\end{align}
since the Hartree contribution vanishes in this approximation. Here, $\beta=1/k_BT$ defines the thermal energy in terms of the Boltzmann constant $k_B$ and temperature $T$.

While the equation for the Green function should be solved self-consistently, for the present purposes it is sufficient to replace the propagators in the self-energy with their corresponding bare ones, $\bfg_\text{mol}(z)=(z-\bfepsilon)^{-1}$ and $D_\bfq(z)=2\omega_\bfq/(z^2-\omega_\bfq^2)$. We, then, write the self-energy as $\tilde\bfSigma_\bfq=\tilde\Sigma_{0\bfq}\sigma_0+\tilde\bfSigma_{1\bfq}\cdot\bfsigma$ where
\begin{subequations}
\begin{align}
\tilde\Sigma_{0\bfq}(z)=&
	\frac{1}{2}
	\sum_{s=\pm1}
	\biggl(
		\frac{1 - f(\dote{s}) + n_B(\omega_\bfq)}
			{z -\dote{s}-\omega_\bfq}
			+
		\frac{ f(\dote{s}) + n_B(\omega_\bfq)}
			{z -\dote{s}+\omega_\bfq}
	\biggr)
\\
\tilde\bfSigma_{1\bfq}(z)=&
	\frac{\hat\bfepsilon_1}{2}
	\sum_{s=\pm1}
	s
	\biggl(
		\frac{1-f(\dote{s}) + n_B(\omega_\bfq)}{z -\dote{s}-\omega_\bfq}
		+
		\frac{f(\dote{s})+n_B(\omega_\bfq)}{z-\dote{s}+\omega_\bfq}
	\biggr)
\end{align}
\end{subequations}
where $\dote{1}=|\boldsymbol{\epsilon}_1|$, $\dote{s}=\dote{0}+s\dote{1}$, and $\hat{\epsilon}_1=\boldsymbol{\epsilon}_1/\dote{1}$, whereas $ n_B(\omega)$ denotes the Bose-Einstein distribution function.

In the following, our aim is to emphasize how the thermal reservoir influences the temperature dependency of the induced magnetic moment. Therefore, we investigate a molecule with unpolarized level, that is, setting $\bfepsilon_1=0$.

In this limit, the unperturbed Green function simplifies to $\bfg(z)=\sigma_0/(z-\dote{0})$ and the electron energy $\dote{s}\rightarrow\dote{0}$ in the self-energy, as well as $\tilde\bfSigma_1\rightarrow0$. Nevertheless, because of the form of the electron-phonon coupling it can be seen that $\Sigma_0=\sum_\bfq(u_{0\bfq}u_{0\bar\bfq}+\bfu_{1\bfq}\cdot\bfu_{1\bar\bfq})\tilde\Sigma_{0\bfq}$ while $\bfSigma_1=\sum_\bfq(u_{0\bfq}\bfu_{1\bar\bfq}+\bfu_{1\bfq}u_{0\bar\bfq}+i\bfu_{1\bfq}\times\bfu_{1\bar\bfq})\tilde\Sigma_{0\bfq}$.  Then, for $\bfG_\text{mol}^r=G^r_0\sigma_0+\bfG^r_1\cdot\bfsigma$, we have
\begin{subequations}
\label{eq-G01}
\begin{align}
\label{G0}
G^r_0(\omega)=&
	\frac{z-\dote{0}-\Sigma^r_0}{(z-\dote{0}-\Sigma^r_0)^2-\bfSigma^r_1\cdot\bfSigma^r_1}
	,
\\
\label{G1}
\bfG^r_1(\omega)=&
	\frac{\bfSigma^r_1}{(z-\dote{0}-\Sigma^r_0)^2-\bfSigma^r_1\cdot\bfSigma^r_1}
	.
\end{align}
\end{subequations}

\begin{figure}[t]
\begin{center}
\includegraphics[width=\columnwidth]{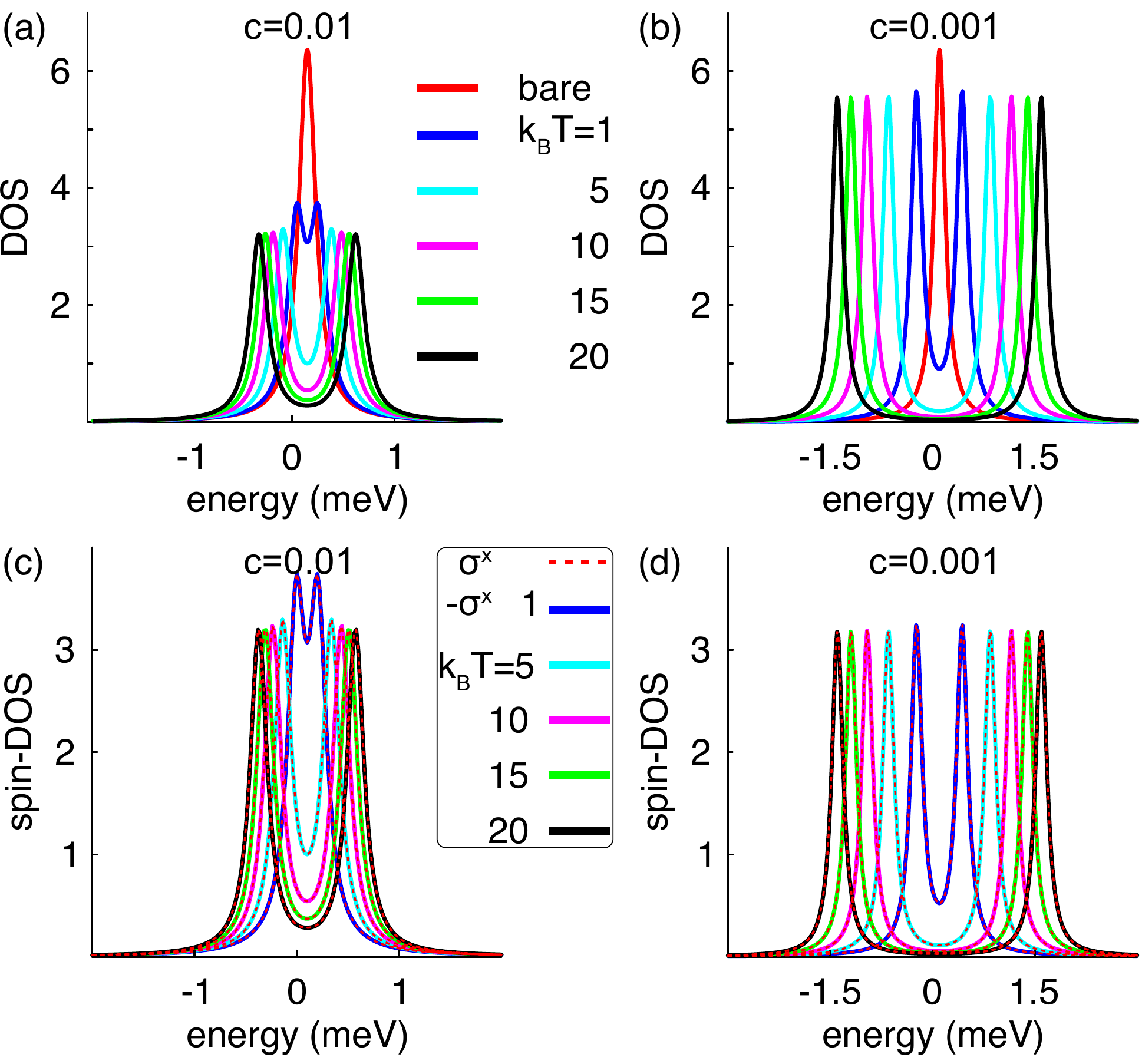}
\end{center}
\caption{(a), (b) Local DOS and (c), (d) local spin-DOS as a function of the energy $\omega$, for the set-up $\dote{0}=0.1$, $u_{0\bfq}=0.01$, $\bfu_{1\bfq}=0$, and phonon velocity (a), (c), $c=0.01$ and (b), (d), $c=0.001$, for temperatures corresponding to the energies $1/\beta\in\{1,\ 5,\ 10,\ 15,\ 20\}$ [units: meV]. The unperturbed (bare) DOS is shown for reference (red).}
\label{fig-Fig1}
\end{figure}

First, we notice in this limit that, a configuration such that $u_{0\bfq}=0$ and $\bfu_{1\bfq}\neq0$, may lead to a modification of the electronic state. The requirement is that $\bfu_{1\bfq}\times\bfu_{1\bar\bfq}\neq0$. The momentum dependence of the coupling rate $\bfu_{1\bfq}$ is related to the phononic polarization vector $\boldsymbol{\epsilon}_\bfq$ which, in turn, depends on the lattice symmetries. For instance, inversion symmetry implies that $\boldsymbol{\epsilon}^*_\bfq=\boldsymbol{\epsilon}_{\bar\bfq}=\boldsymbol{\epsilon}_\bfq$, under which conditions, then, the self-energy $\bfSigma_1=0$, hence, also $\bfG_1=0$. On this note, it is relevant to mention that chiral phonons, for which there is no inversion symmetry, would open for the possibility to generate an electronic spin-polarization, something that was considered in Ref. \citenum{arxiv.2210.12722}.

From the expressions in Eq. \eqref{eq-G01}, we calculate the local density of electron states $\av{n_\text{mol}}=(-i){\rm sp}\int\bfG^<_\text{mol}(\omega)d\omega/2\pi=(-i)\int G^<_0(\omega)d\omega/\pi$, which, for $u_{0\bfq}=u_0=0.01$ and $\bfu_{1\bfq}=0$, is plotted in Fig. \ref{fig-Fig1} (a), (b), as a function of the energy for temperatures corresponding to thermal energies between 1 meV and 20 meV. The unperturbed (bare) density of states has a single peak at the energy $\dote{0}=0.1$ (red). When the electron-phonon interaction is turned on, this central peak splits into two which are located symmetrically around $\dote{0}$. This is expected considering the poles given in Eq. \eqref{eq-poles}. 
The plots in Fig.  \ref{fig-Fig1} (a), (b), illustrate the thermal evolution of the density of state for two different phonon velocities, (a) $c=0.01$ and (b) $c=0.001$. The width of the spectrum is expected to increase inversely with the velocity, since more phonon modes contribute to the interactions with the electron the lower the velocity.

\begin{figure}[t]
\begin{center}
\includegraphics[width=\columnwidth]{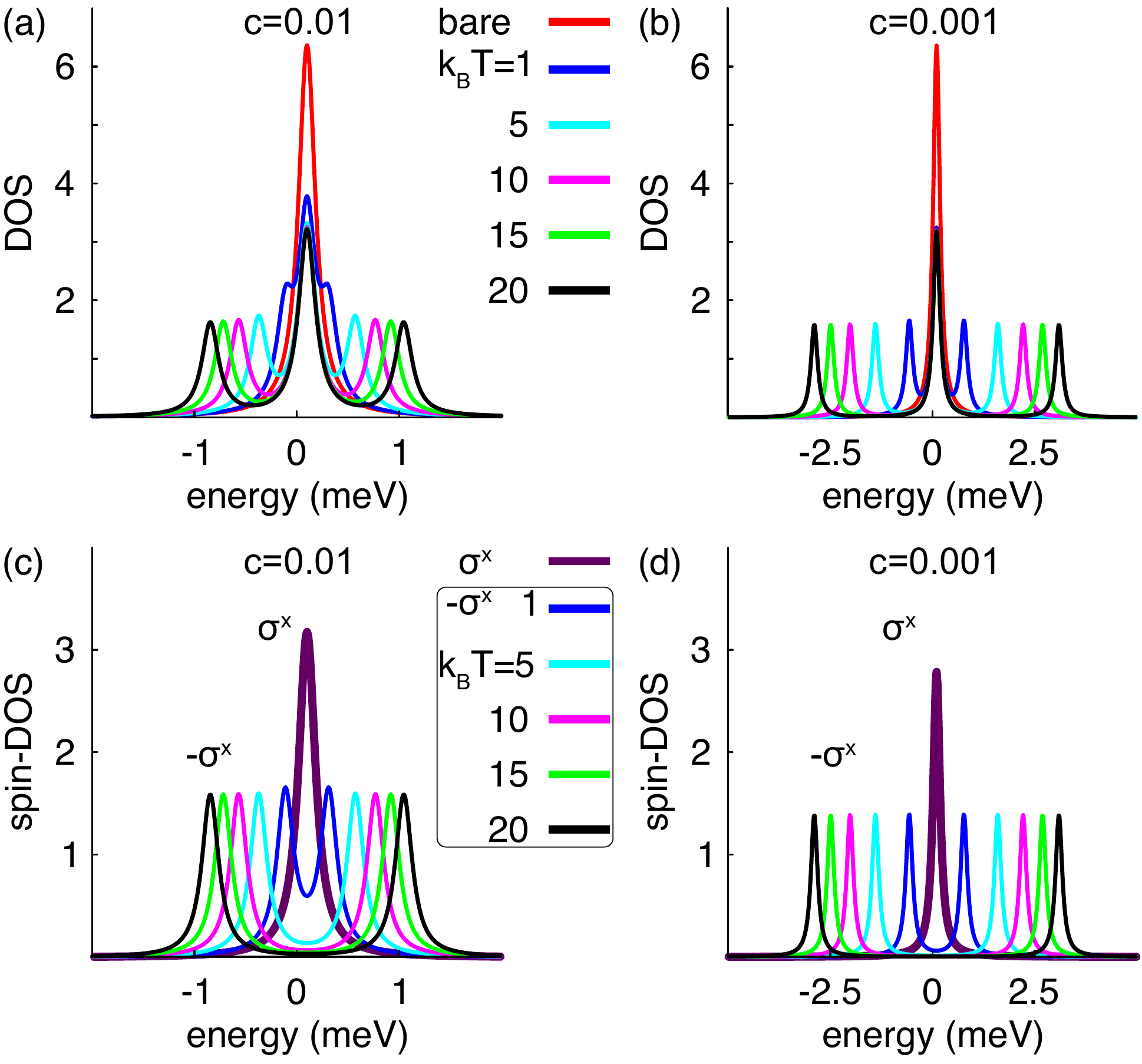}
\end{center}
\caption{(a), (b) Local DOS and (c), (d) local spin-DOS as a function of the energy $\omega$, for the set-up $\dote{0}=0.1$, $u_{0\bfq}=0.01$, $\bfu_{1\bfq}=0.01\hat{\bf x}$, and phonon velocity (a), (c), $c=0.01$ and (b), (d), $c=0.001$, for temperatures corresponding to the energies $1/\beta\in\{1,\ 5,\ 10,\ 15,\ 20\}$ [units: meV]. The unperturbed (bare) DOS is shown for reference (red).
}
\label{fig-Fig2-3}
\end{figure}

Despite the splitting of the density of electron states, the spin degeneracy remains preserved. This is clear since the electron-phonon coupling only contains the spin-conserving component. This leads, trivially, to that $\bfSigma_1=0$, hence, the spin-dependent component $\bfG_1$ of the Green function also vanishes. For completeness, the spin-resolved density of electron states are plotted in Fig. \ref{fig-Fig1} (c), (d), illustrating the degeneracy of the spin projections.

The combination of charge and spin coupling interactions with the phonons, on the other hand, results in the emergence of two resonance peaks alongside the initial elastic peak in the density of state. This is illustrated in Fig. \ref{fig-Fig2-3} for $\bfu_{1\bfq}=u_0\hat{\bf x}$ and otherwise the same conditions as for the plots in Fig. \ref{fig-Fig1}. The side peaks shift to the higher energies with increasing temperature, while the central peak acquires a lowered amplitude. Also here, the lower velocity tends to induce a stronger shift of the side peaks with increasing temperature, as expected from the previous case.

\begin{figure}[t]
\begin{center}
\includegraphics[width=\columnwidth]{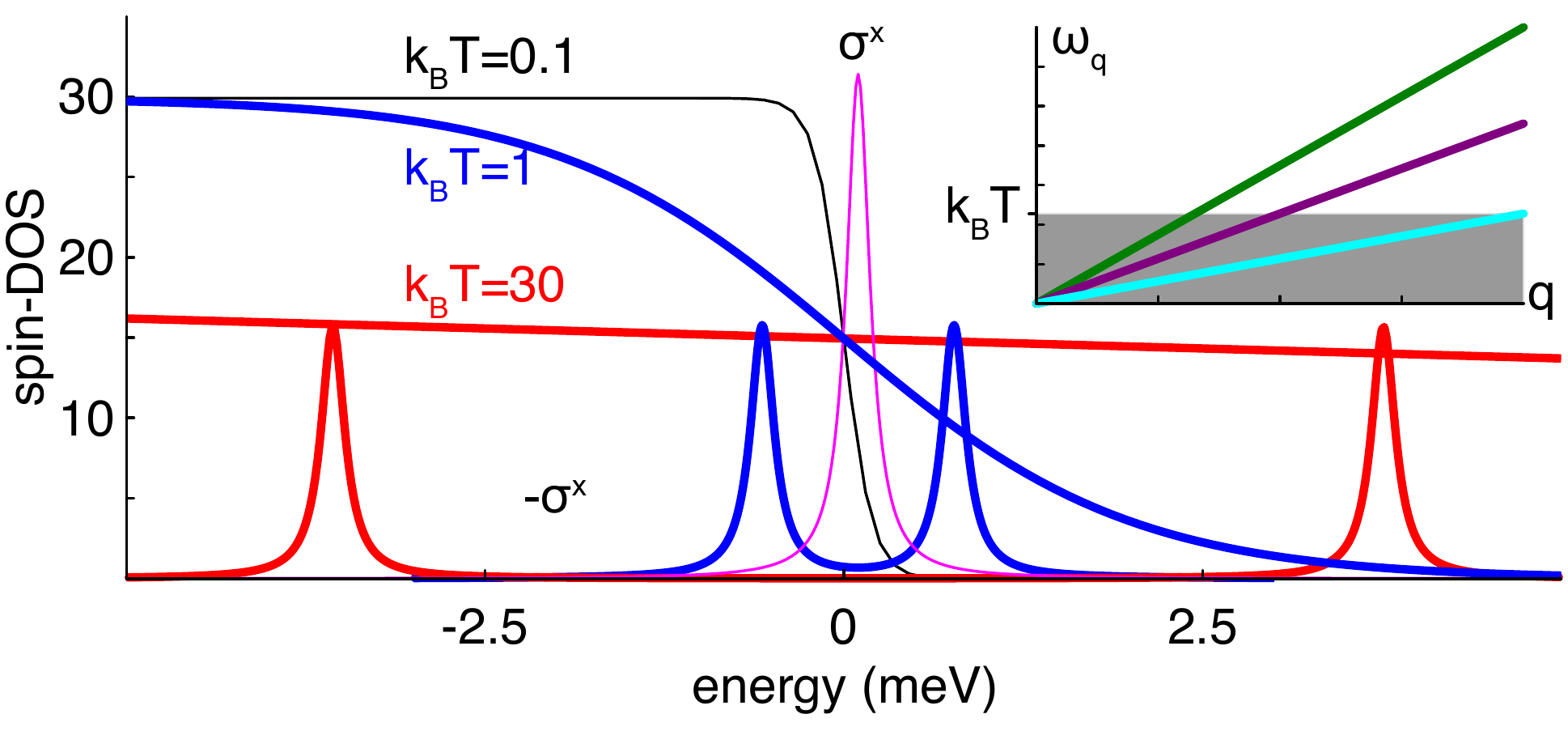}
\end{center}
\caption{Influence of the temperature on the magnetic moment, both with respect to the shifts of the inelastic resonances and the thermal occupation factor (Fermi-Dirac distribution function). At low temperatures (blue), the inelastic resonances are strongly asymmetrically occupied while the occupation become symmetrized with increasing temperature (red). Very low temperatures is shown for reference (black). The spin-resolved densities are given for the conditions in Fig. \ref{fig-Fig2-3} (d).
The inset illustrates the phonon dispersions for three different velocities, (cyan) low, (purple) moderate, and (green) high, and the expected thermally occupied states (gray area) for each dispersion relation.}
\label{fig-ResAndFermi}
\end{figure}

In order to draw any conclusions about the spin properties under these conditions, however, we investigate the spin resolved densities of states captured in the matrix $\bfrho(\omega)=-\im\bfG^r_\text{mol}(\omega)/\pi$. The spin resolved densities of states are plotted in Fig. \ref{fig-Fig2-3} (c), (d). As expected, the spin-dependent coupling $\bfu_{1\bfq}u_0\hat{\bf x}$ breaks the degeneracy of the electronic structure. Quite unexpectedly at first glance, on the other hand, is that the spin projections are separated into two mutually exclusive branches.
Here, however, this is not surprising since the self-energies $\Sigma^r_0$ and $\bfSigma^r_1$ are both proportional to $\tilde\Sigma^r_0$ and $\bfu_{1\bfq}=u_0\hat{\bf x}$, which leads to that $\bfG^r_\text{mol}$ can be partitioned into
\begin{align}
\bfG^r_\text{mol}=&
	\frac{1}{2}
	\frac{\sigma^0+\sigma^x}{\omega-\dote{0}+i\delta}
	+
	\frac{1}{2}
	\frac{\sigma^0-\sigma^x}{\omega-\dote{0}-4u_0^2\tilde\Sigma^r_0}
	,
\end{align}
where $\delta>0$ infinitesimal.

This partitioning makes it clear that one central resonance is located at the elastic energy $\omega=\dote{0}$, whereas the other resonances are found at the condition $\omega-\dote{0}-4u_0^2\tilde\Sigma_0^r=0$, an equation which in the current approximation has two solutions. In Fig. \ref{fig-Fig2-3} (c), (d), the resonances corresponding to the first and second contributions are signified by $\pm\sigma^x$.

\begin{figure}[t]
\begin{center}
\includegraphics[width=\columnwidth]{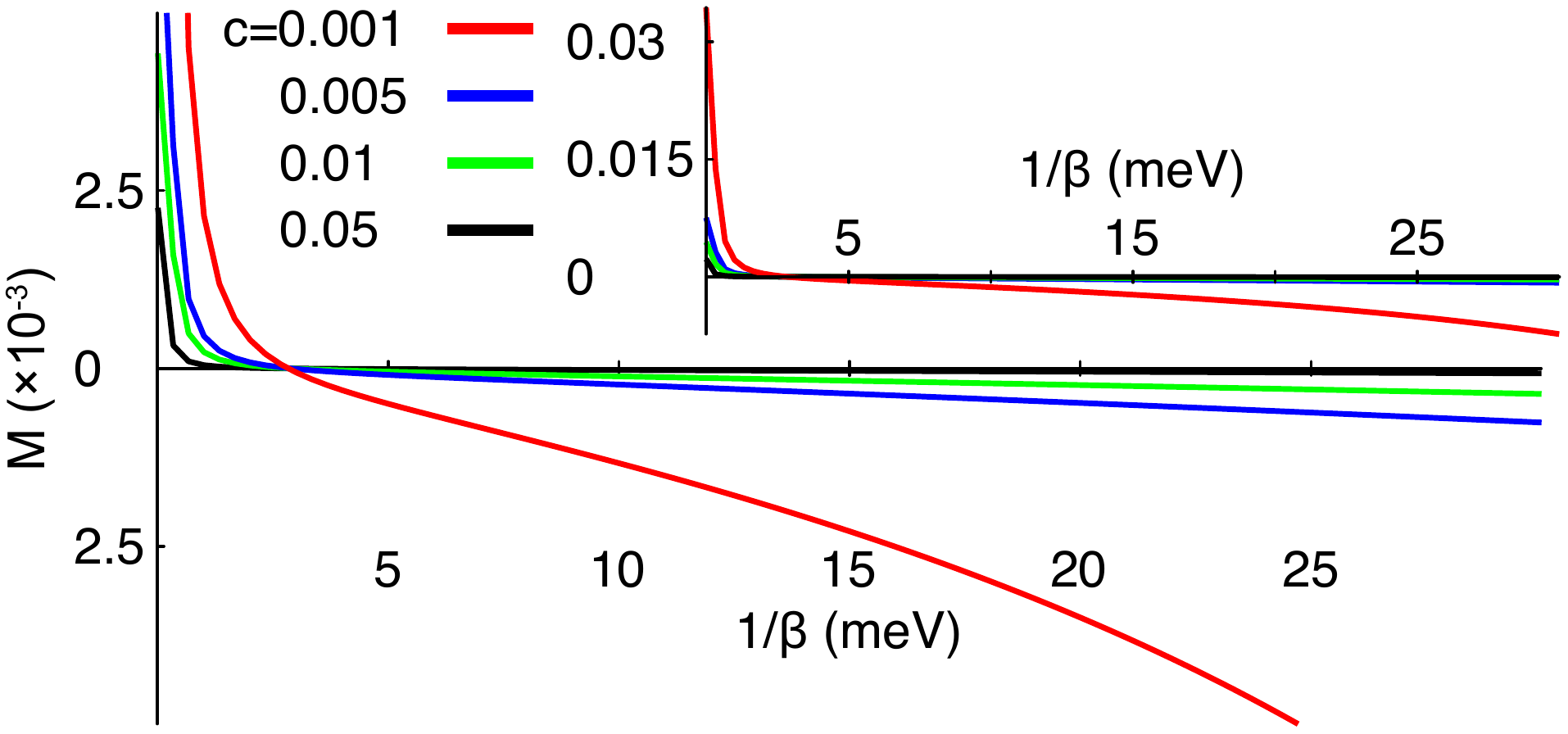}
\end{center}
\caption{Induced magnet moment as a function of the thermal energy $1/\beta$, for the set-up $\dote{0}=0.1$, $u_{0\bfq}=0.01$, $\bfu_{1\bfq}=0.01\hat{\bf x}$, and phonon velocity (blue) $c=0.001$, (red) $c=0.005$, (black) $c=0.01$, and (green) $c=0.05$ [units: meV].}
\label{fig-Fig4}
\end{figure}

The associated molecular magnetic moment $\av{\bfM_\text{mol}}=\calM\hat{\bf x}$ resulting from these conditions is given by
\begin{align}
\calM=&
	\frac{1}{2}
	f(\dote{0})
	+
	\im
	\int
		\frac{f(\omega)}{\omega-\dote{0}-4u_0^2\tilde\Sigma^r_0}
	\frac{d\omega}{4\pi}
	.
\label{eq-Mmolx}
\end{align}
While this moment is, in general, non-vanishing, it undergoes a sign change at a finite temperature $T_\text{xo}$. This is understood by the opposite signs of the two contributions constituting $\calM$ in Eq. \eqref{eq-Mmolx}; recall that $\im(\omega-\dote{0}-4u_0^2\tilde\Sigma_0^r)^{-1}<0$. Here, the first contribution, which is positive, dominates the magnetic moment at low temperature, see Fig. \ref{fig-ResAndFermi}. Put simply, in the figure it can be seen that whereas the central resonance at $\omega=\dote{0}$ is nearly fully occupied, the side resonances are only partially occupied. Since the former and latter resonances add positively and negatively, respectively, to the total moment, the moment is positive at sufficiently low temperature. This property is corroborated by our computations of the magnetic moment, see Fig. \ref{fig-Fig4}, which displays $\calM$ as a function of the temperature for different phonon velocities $c$.

With increasing temperature, the occupations of all resonances increase, however, while the occupation of the central resonance is marginally increased, the side resonances approach full occupation such that the two branches cancel out each other. Nevertheless, since the overall spin-density has a slight overweight at the side resonances, this contribution eventually becomes larger than the central resonance such that the total moment changes sign at a cross-over temperature $T_\text{xo}$ and becomes negative. The sign change and negative moment is clearly illustrated in Fig. \ref{fig-Fig4}, which also shows that the amplitude, both positive and negative, of the moment increases with decreasing velocity $c$. 

The latter observation is understood in terms of the thermally accessible energies for a given temperature, see inset of Fig. \ref{fig-ResAndFermi}, illustrating the phonon dispersion relations for three velocities, (blue) low, (red) moderate, and (black) high, and the expected thermally occupied states (gray area) for each dispersion relation. For phonons with low velocity, a lower temperature is required to thermally access the energies for a larger portion of the phononic $\bfq$-vectors in reciprocal space, compared to phonons with a higher velocity. Therefore, it is not surprising that slow phonons contribute more to the limiting magnetic moments, than fast phonons.

\begin{figure}[t]
\begin{center}
\includegraphics[width=\columnwidth]{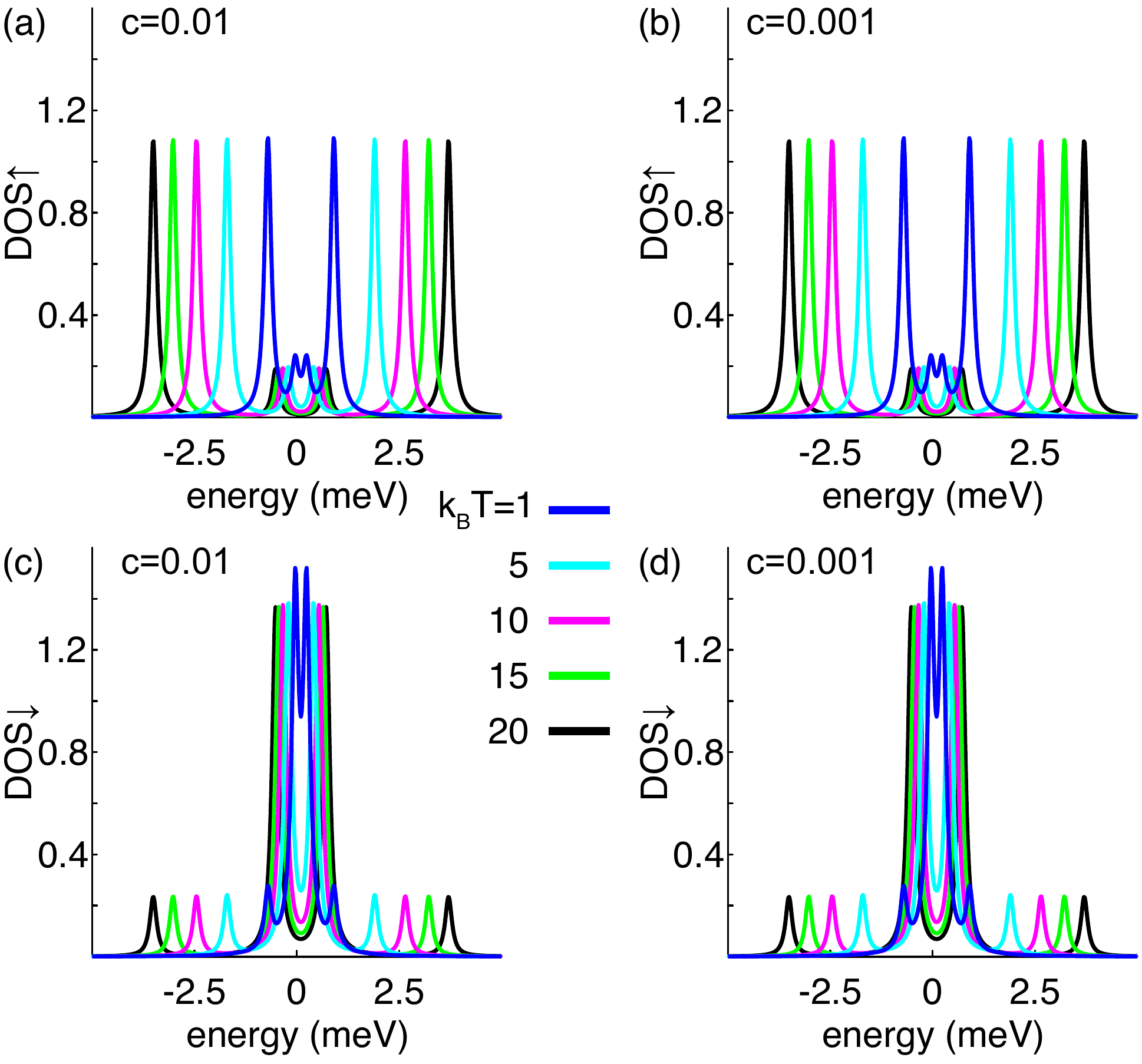}
\end{center}
\caption{Local spin-DOS with (a), (b), spin $\up$-projections and (c), (d), spin $\down$-projection, as a function of the energy $\omega$, for the set-up $\dote{0}=0.1$, $u_{0\bfq}=0.01$, $\bfu_{1\bfq}=0.01(\hat{\bf x}+\hat{\bf z})$, and phonon velocity (a), (c) $c=0.01$ and (b), (d) $c=0.001$, for temperatures corresponding to the energies $1/\beta\in\{1,\ 5,\ 10,\ 30\}$ [units: meV].}
\label{fig-Fig5}
\end{figure}

Finally, we consider the configuration with $\bfu_{1\bfq}=u_0(1,0,1)$. For these conditions, the molecular Green function can be written
\begin{align}
\bfG^r_\text{mol}=&
	\frac{1}{4}
	\sum_{s=\pm1}
	\frac{2\sigma^0+s\sqrt{2}(\sigma^x+\sigma^z)}{\omega-\dote{0}-(3+s2\sqrt{2})u^2_0\tilde\Sigma^r_0}
	.
\end{align}
In this set-up, there is no clear separation of the central and side resonances, instead the two branches mix. In Fig. \ref{fig-Fig5}, we display plots of the spin resolved density of electron states, for the same conditions as in Fig. \ref{fig-Fig2-3}, however, with $u_{z\bfq}\neq0$. First, one may notice that the resonances are mixtures of both spin projections. Second, it is clear that one spin branch is more heavily weighted on the side resonances, Fig. \ref{fig-Fig5} (a), (c),  whereas the other branch has an overweight on the central resonance, Fig. \ref{fig-Fig5} (b), (d), albeit the central resonance cannot be clearly resolved.

In fact, the central resonance cannot be identified as a single resonance under the given conditions, since the electronic density comprises four distinct peaks. The four resonances can be found as the solutions to real parts of the two equations $\omega-\dote{0}-(3\pm2\sqrt{2})u_0^2\tilde\Sigma^r_0=0$, of which the $+$ ($-$) equation provides the resonances which are more heavily weighted on the side (central) resonances. In this sense, each equation corresponds to one of the two spin branches and despite the mixing between these, one can identify a slight discrimination between them.

\begin{figure}[t]
\begin{center}
\includegraphics[width=\columnwidth]{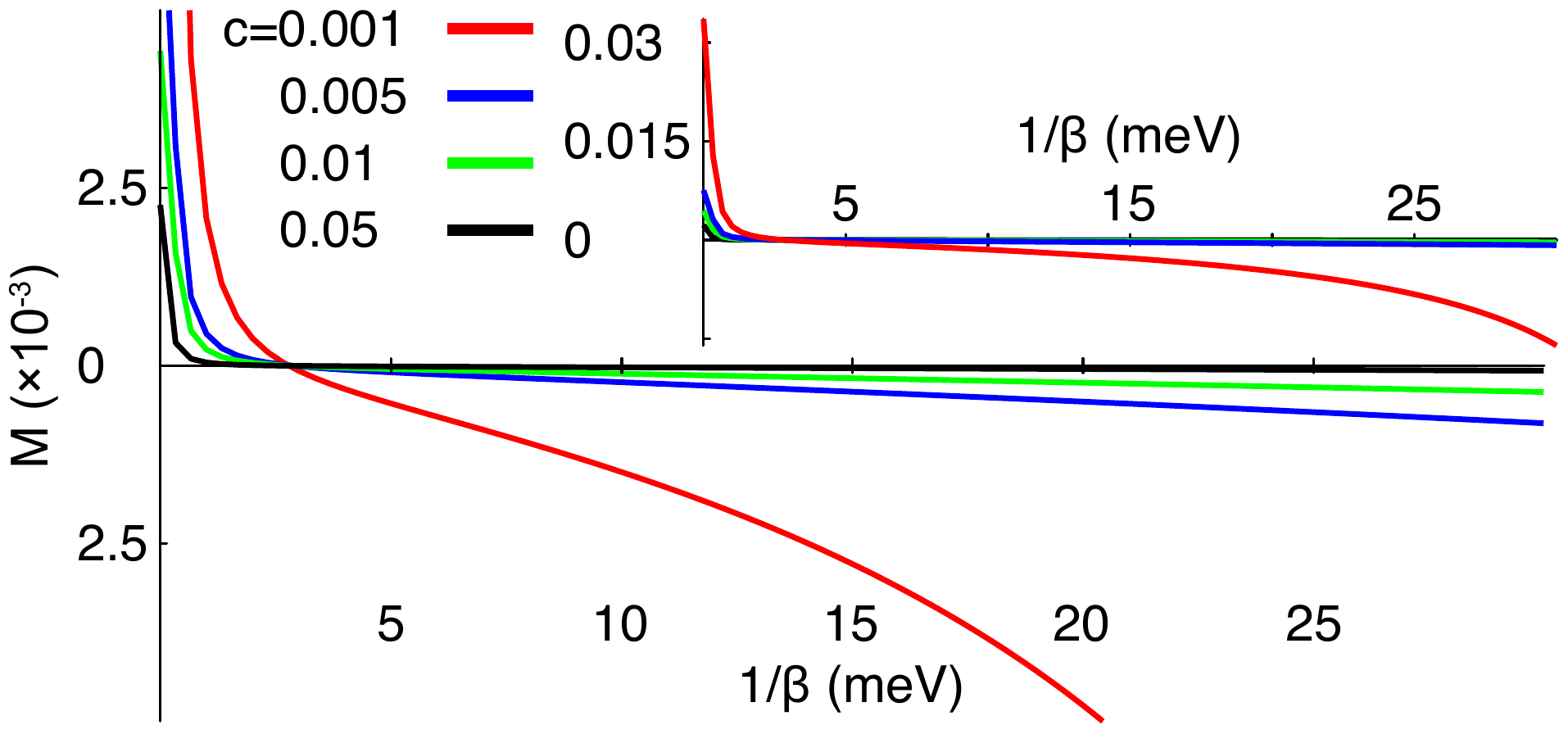}
\end{center}
\caption{Induced magnet moment as a function of the thermal energy $1/\beta$, for the set-up $\dote{0}=0.1$, $u_{0\bfq}=0.01$, $\bfu_{1\bfq}=0.01(\hat{\bf x}+\hat{\bf z})$, and phonon velocity (blue) $c=0.001$, (red) $c=0.005$, (black) $c=0.01$, and (green) $c=0.05$ [units: meV].}
\label{fig-Fig6}
\end{figure}

In this configuration, the induced molecular magnetic moment can be written $\av{\bfM_\text{mol}}=\calM(\hat{\bf x}+\hat{\bf z})$, where the factor $\calM$ is provided by the integral
\begin{align}
\calM=&
	\frac{\sqrt{2}}{4}
	\sum_{s=\pm1}
	\int
		\frac{sf(\omega)}{\omega-\dote{0}-(3+s2\sqrt{2})u^2_0\tilde\Sigma^r_0}
	\frac{d\omega}{2\pi}
	.
\label{eq-Mxz}
\end{align}
Again, we can identify a cross-over temperature $T_\text{xo}$ at which the total moment changes sign from positive to negative, which can be seen in Fig. \ref{fig-Fig6}, in which the factor $\calM$ is plotted as a function of the temperature, for different phonon velocities. The mechanism for this sign change is the same as in the previous configuration. Whereas one spin-projection becomes more or less fully occupied already at low temperature and the other is only partially occupied, the latter tends to become increasingly occupied with the temperature and eventually dominates the overall magnetic moment. It can also be observed that the total magnetic moment increases when the $z$-component is added to the already existing $x$-component of the interaction parameter $\bfu_{1\bfq}$. This observation is, however, trivial due to the increased number of scattering channels that are opened.

A more important, and also interesting observation that may be done, is that the temperature for the sign change of the magnetic moment appears to be universal and independent of the phonon velocity, see Figs. \ref{fig-Fig4}, \ref{fig-Fig6}. This property is not surprising when considering that the sign change is a result of the competition between the two contributions, c.f., Eqs. \eqref{eq-Mmolx}, \eqref{eq-Mxz}. The two contributions have equal temperature dependencies irrespective of the phonon velocity which, therefore, leads to that the specific phonon distribution does not impact the temperature at which the two contributions cancel. Should the two contributions, on the other, have unequal dependencies on the phonon distribution, then the cross-over temperature may vary with, e.g., the phonon velocity. Currently, we are not aware of which type of electron-phonon interactions that would cause such inhomogenous temperature dependencies, however, it is possible that structures in which the phonons modes are strongly anisotropic would open up for such properties.


In summary, we have theoretically investigated the influence of combined interactions on a localized level and demonstrated that an electronic state may become spin-polarized when coupled to reservoirs. We show that a system which is non-magnetic whenever isolated from an surrounding environment, may spin-polarize when a connection to such environment is made. The system may spin-polarize if there are, at least, two types of interactions of which at least one has an intrinsic spin-dependence associated with it. Formally, an interaction that can be expressed as $V_0\sigma^0$ and $\bfV_1\cdot\bfsigma$, changes the electronic spectrum by $(V_0\sigma^0+\bfV_1\cdot\bfsigma)^2=(V_0^2+|\bfV_1|^2)\sigma^0+2V_0\bfV_1\cdot\bfsigma$. Hence, the electronic spectrum becomes spin-dependent if and only if both $V_0$ and $\bfV_1$ are non-zero. Under those conditions, there is a potential for the system to acquire a non-vanishing magnetic moment.

As a corollary, we develop a theory for temperature-dependent magnetization in a molecule. We show that spin-dependent inelastic scattering, e.g., off phonons which may arise due to spin-orbit coupling \cite{PhysRevB.102.235416}, leads to breaking of the time-reversal symmetry. For this, we employ an unconventional treatment of electron scatterings off phonons by taking into account both the charge-phonon and spin-phonon couplings. While none of these coupling individually break the electronic spin degeneracy, our findings show that the combination of the two leads to a splitting of the spin channels. The effect we consider, which results in non-conserved energy collisions, originates from the interplay between spin-orbit coupling and vibrational modes. We, furthermore, demonstrate that the inelastic scattering does induce a non-zero magnetic moment of the initially unpolarized molecule, a moment which magnitude increases with temperature, however, changes sign at a cross-over temperature. The sign change of the magnetic moment can be explained in terms of competing influences from the relevant interactions.

Despite that we are currently aware of experimental results which comply with our theoretical discussion \cite{NatComms.8.14567,Small.1801249,JPhysChemLett.10.1139,PhysRevMaterials.5.114801,ChemSci.13.12208,JACS.144.7302}, it would intriguing to consider the effects under more extreme conditions, for instance, measurements of magnetically asymmetric thermopower or using magnetic force microscopy to measure asymmetric forces of chiral molecules attached to surface.


\bibliography{ref}

\end{document}